\documentclass{article}
\usepackage{authblk}
\usepackage[super]{cite}
\usepackage{xcolor}
\usepackage[
    starfontserif 
    ]{starfont}
\usepackage[verbose,hypertexnames=false]{hyperref}
\usepackage{latexsym,amssymb,amsmath,bbm,epsf}
\usepackage{float}
\usepackage{graphicx} 
\usepackage{xcolor} 
\usepackage{cite,comment}
\hypersetup{colorlinks=false,allbordercolors=blue,pdfborderstyle={/S/U/W 1}}
\providecommand{\keywords}[1]
{
  \small	
  \textbf{\textit{Keywords---}} #1
}

\begin{document}

\markboth{Foerster \& LeBohec}{Scale Relativity signatures in extra-solar planetary systems}

\title{Resolution-Scale Relativity signatures in the orbital periods of extra-solar planetary systems}

\author{Julien Foerster
}

\affil{\small Ecole du Bois Sauvage, Rue Van Aa 10, \\1050 Bruxelles, Belgium\\
jfoerster.private@gmail.com}

\author{Tugdual LeBohec}

\affil{\small University of Utah, Department of Physics and Astronomy, 115 South 1400 East
\\Salt-Lake-City, UT 84112-0830, USA\\
lebohec@physics.utah.edu}

\maketitle

\begin{abstract}
    Resolution-Scale Relativity suggests quantum-like dynamics may emerge in chaotic macroscopic systems. In planetary systems, this would lead to orbital periods being proportional to cubed integers $n$. Each system is then characterized by a fundamental speed corresponding to orbital $n=1$. Fitting this model to data from the NASA Exoplanet Archive for 115 planetary systems with four or more planets leads to identifying 38 systems (33\%) complying with an accuracy such that the null hypothesis accidental probability is less than $10^{-2}$, and 16 (14\%) with less than $10^{-3}$. Additionally, 34 systems (29\%) follow a pattern of consecutive quantum-like integer numbers, and 101 (88\%) in which at least half of the quantum-like numbers are part of consecutive sequences. The distribution of fundamental speeds extends from $\sim 100\,\rm km/s$ to more than $1,200\,\rm km/s$ and can be described in terms of a few peaks centered on integer multiple of a super-fundamental speed $v_0=(218.0\pm4.7)\,\rm km/s$.  These results along side with other observations in turbulent fluid dynamics amount to a shift to a higher gear in the search for macro-quantization effects. 
    \end{abstract}

\keywords{Resolution-Scale Relativity; Macro-quantization; Extrasolar planetary systems.}

\section{Introduction}\label{intro}
Less than twenty years after the formulation of the Rutherford-Bohr model of the atom \cite{bohr1913}, and just a few year after the publication of Schr\"odinger's equation \cite{schrodinger1926}, a few physicists and astronomers reported their observation that the Solar System and the systems of moons around some of the major planets are structured in a way that is reminiscent of the orbitals in a {\it hydrogen-like atom} with orbital semi-major axis quite precisely scaling in proportion to squared integers \cite{caswell1929,malisoff1929,peniston1930}.

There was no theoretical framework justifying any similarity between gravitational Keplerian systems and atoms until possibly the development of stochastic mechanics by E.~Nelson \cite{Nelson1966} who showed that, for a particle of mass $m$, the hypothesis of a sub-quantum Brownian dynamics characterized by a diffusion constant $\mathcal D$ with the influence of external fields accounted for by means of Newton's fundamental relation of dynamics naturally led to Schr\"odinger's equation with Planck's constant $\hbar$ replaced with $2m\mathcal D$. The fundamental nature of this Brownian motion was unspecified. Although it was not explored at the time, this opens up on the possibility of a transposition to macroscopic system with the effectively stochastic dynamics of chaotic systems playing the role of Nelson's sub-quantum Brownian motion. 

A more consistent theoretical framework was then presented by L.~Nottale \cite{nottale1993,nottale2011} who considered the extension of the principle of relativity to changes of resolution-scales. Reference frames are usually characterized by their relative positions, orientations and motions. The principle of relativity then requires the fundamental laws of nature to retain the same form when expressed in different reference frames. Nottale's {\it Resolution-Scale Relativity} amounts to having reference frames being also specified by their relative resolution-scales and to extend the relativity principle to changes in resolution-scales. This does not affect Newtonian dynamics, as refinements of the resolution-scale used to describe trajectories only improve the precision without revealing new or different structures or properties. The situation is  however different in the quantum domain and with chaotic/complex or stochastic systems as, with them, the relevance of the concept of trajectory vanishes. The refinement of the resolution-scale does not improve the precision with which a trajectory would be described, but instead it reveals new features in the motion and, sometimes, entirely new aspects of the system.  This amounts to incorporating fractal and therefore non-differentiable dynamical paths in the evolution of the system. Consequently, the implementation of resolution-scale relativity shares some technical aspects with Nelson's stochastic quantization but the interpretation is entirely different. In particular, instead of resting on an additional hypothesis (sub-quantum Brownian motion), the resolution-scale relativity approach proceeds by relaxing the unnecessary assumption of differentiability. An attempt at measuring the trajectory of the system selects a bundle of indistinguishable and non-differentiable dynamical paths. This bundle specifies the state of the system without any one path actually being followed \cite{mhteh2018}. In particular, the quantum or {\it quantum-like} dynamics only depends on the property of the non-differentiable dynamical paths under infinitesimal transformations of the resolution-scale. Specifically, the {\it quantum-like} dynamics does not depend on the non-differentiable character of the dynamical paths to be preserved all the way down to infinitesimal resolution-scales. This opens the possibility for {\it quantum-like} dynamics emerging from the Resolution-Scale Relativity principle to be applicable to macroscopic systems, whose chaotic or complex nature maybe described in terms of stochastic paths over some finite range of resolution-scales. In this framework, the enforcement of the Resolution-Scale Relativity principle in the construction of a theory of point dynamics  with time as an external and absolute parameter leads to a natural and consistent foundation of quantum mechanics \cite{nottale1993,nottale2011, prodanov2021}.

This motivates the search for {\it quantum-like} dynamics signatures in macroscopic systems. Such an observation would indicate that the resolution-scale extension of the relativity principle is actually implemented in nature. Ideally, this type of Resolution-Scale Relativity signatures would be observed in the reproducible environment of the laboratory. Unfortunately, it seems difficult to identify laboratory setups that are both non-dissipative and sufficiently simple from the point of view of the number of forces at play to be simply and accurately tractable with the methods of quantum mechanics. Proposals have been formulated to search for {\it quantum-like} signatures in externally maintained turbulent hydrodynamic flows \cite{nottale2011} or in the Brownian motion of a micro-sphere in an optical trap \cite{lebohec2017}.  Alternatively, one can search for such signatures in astrophysical systems such as planetary systems. The 1929-30 identifications of quantum-like structures in the solar system by Caswell, Malisoff, and Penniston \cite{caswell1929,malisoff1929,peniston1930} were revisited in more detail in the Resolution-Scale Relativistic context \cite{nottale1997,hermann1997} and even included an account for the masses of the major objects of the Solar system following a {\it quantum-like} hydrogen orbital profile. Similar analyses were performed for Kuiper belt objects in the Solar System \cite{nottale2011}, extra-solar planetary systems\cite{nottale2000}, binary stars, pairs of galaxies, and others \cite{nottale2011}. 
The conclusions of these studies are all indicative or suggestive of a {\it quantum-like} structuring in these otherwise classical gravitational systems. Predictions of macro-quantization in extra-solar planetary systems preceded the onset of their discoveries \cite{nottale1993} and motivated early studies \cite{nottale1996}. In the present article, after using the solar system as a prototype for our analysis and reproduce some results from earlier studies\cite{hermann1997,nottale1997}, we analyze recent data from observations of extra-solar planetary systems.

The paper is organized as follows. In Section \ref{analysis}, we review the transposition of the quantum hydrogen atom model to classical gravitational Keplerian systems substituting $\hbar$ with $2m\mathcal D$ while replacing the coulombic potential with the gravitational potential. The relation between orbital periods and principal quantum-like numbers suggests how the data should be analyzed but principal quantum-like numbers need to be assigned to the different planets within each planetary system. We present the method used for this in Subsection \ref{nassignalgo}.  The analysis is exemplified in Section \ref{solar} with applications to the Solar System.
Then, Section \ref{fiveandmoreplntssyst} turns to the analysis of extra-solar planetary systems with four or more detected planets.  The systematic effects that could result from orbital resonances are discussed in Section \ref{resonance}. Finally, Section \ref{closing} recapitulates the findings and discusses their possible implications.

\section{Analysis}\label{analysis}
\subsection {The macro-quantum gravitational hydrogen model \label{bohr}}
If the Resolution-Scale Relativity principle is implemented in nature in a way that results in macroscopic quantization, we can expect objects in Keplerian potentials to be distributed in a way that corresponds to the solutions of Schr\"odinger's equation  with $\hbar$ replaced with $2m\mathcal D$, in which $\mathcal D$ is a diffusion coefficient :
\begin{equation}
    i2m\mathcal{D}\frac{\partial \psi}{\partial t}=-2m\mathcal{D}^2\nabla^2 \psi - \frac{GMm}{r}\psi.
\end{equation}
 In this equation, $G$ is the universal constant or gravitation, and $M$ is the mass of the central object. The mass $m$ of the orbiting object is considered very small compared to that of the central object. Because $m$ appears in the Planck-like constant $2m\mathcal D$, it cancels out from the generalized Schr\"odinger equation. 
This is a consequence of the equivalence principle and corresponds to the fact that the generalized de Broglie wavelength of an object of mass $m$ moving at speed $v$  is $\lambda=\frac{2m\mathcal D}{m v}=\frac{2\mathcal D}{v}$, independently of the mass. This cancellation implies that orbital structuring of gravitational systems is expected to be independent of the masses of individual objects participating in the system's dynamics. Instead of the particle's energy being quantized, it is the mass specific energy, and therefore the speed of the particles, that would be quantized.
 
The stationary solutions are then familiar and expressed in terms of Laguerre polynomials and spherical harmonics \cite{cohentan} with quantized energy eigenvalues 
\begin{equation}
    \frac{E_n}{m} = -\frac{G^2M^2}{8\mathcal{D}^2}\frac{1}{n^2},
\end{equation}
where $n$ is the principal quantum number. The classical relation for the energy of a body having an elliptic trajectory in a Keplerian potential is  
\begin{equation}
    \frac{E_n}{m} = -\frac{GM}{2a_n}= - \Big(\frac{\pi GM}{\sqrt{2}T_n}\Big)^{2/3},
\end{equation}
where $a$ is the semi-major axis of the orbit, and $T$ the orbital period. These two equations imply the quantization of orbital periods following 
\begin{equation}
    T_n = 2\pi \frac{(2\mathcal D)^3}{(GM)^2} n^3 =2\pi{GM}\left(\frac{1}{v_n}\right)^3,\label{bohrt}
\end{equation}
where $v_n$ is the mean orbital speed
\begin{equation}
 v_n = \frac{GM}{2\mathcal D }\frac{1}{n}=\frac{v_F}{n},\label{bohrv}   
\end{equation}
where we introduced the fundamental orbital speed $v_F=\frac{GM}{2\mathcal D}$.

This provides us with an opportunity to search for Resolution-Scale Relativity signatures in the orbital parameters of Keplerian systems, in particular, in extra-solar planetary systems. In the following sections, we will make use of Equation \ref{bohrt} to calculate values of the inverse orbital speed $\eta_n=\frac{1}{v_n}$ from the knowledge of orbital periods $T_n$: 
\begin{eqnarray}
\eta_n&=&\sqrt[3]{\frac{T_n}{2\pi GM}}. \label{etafromt}
\end{eqnarray}
With this, within one planetary system, the proportionality of $\eta_n$ to integers $n$, principal {\it quantum-like} numbers, would be the signature of Resolution-Scale Relativity being at play. With $\eta_F=\frac{1}{v_F}=\frac{2\mathcal D }{GM}$, a system-specific constant, we would have $\eta_n\simeq n \eta_F$. We will use $v_F$ as the single parameter in the tentative {\it quantum-like} description of each planetary system. The different values of $n$ are not known a priori and need to be assigned optimally as described in the next subsection.

In this program, we will use solar system units with the astronomical units $1\,{\rm AU}=149.6\times10^9\,\rm m$ for distances, the year $1\,{\rm yr}=3.156\times10^7\,\rm s$ for times, and the solar mass $M_{\odot}=1.989\times 10^{30}\,\rm kg$ for masses. With these units, the universal constant of gravitation is
$G=4\pi^2\,\rm AU^3M_\odot^{-1}yr^{-2}$, and the mean orbital speed is $v_n=2\pi\sqrt[3]{\frac{M}{T_n}}\,\rm AU/yr$.

Before moving to the data analysis, it is important to recall that other laws have been already proposed to describe the regularity in planets distances to their stars such as Titius-Bodes type scaling laws\cite{Nieto1972}. These are however purely empirical while the relation we propose to test is theoretically founded on a construction of dynamics for non-differentiable paths leading to macro-quantum effects. Furthermore, Titius-Bode laws require more fitting parameters and have difficulty to give a good fit, even for the solar system, without having to remove objects such as Neptune and Pluto and giving Mercury a $-\infty$ position in the sequence. 
\subsection{Principal quantum-like number assignment algorithm}\label{nassignalgo}

While, with $n$ the principal quantum-like number, the relation $\eta_n=\eta_F n$ is best for revealing the quantum-like structure of planetary systems, the equivalent relation $n v_n=v_F$ is more convenient to assign principal quantum-like numbers as follows. For a given system, with letter subscripts here indicating the different planets, we multiply the mean orbital speeds $v_b,\,v_c,\,\cdots$ 
 by a sequence of successive integers ranging from 1 to some large enough value. 
 We then search for the combinations of different orbital speed multipliers $n_b,\, n_c,\, \cdots $ such that $n_b v_b\approx n_cv_c\approx \cdots\approx v_F$. This is illustrated graphically in Figure \ref{assignement}, with the inner solar system as an example.  The optimal fundamental speed and integer multiplier sequence are selected to result in the tightest alignments with the fundamental speed value $v_F$. In practice, this is done by minimizing with respect to $v_F$ the sum of the squared difference between $v_F$ and the closest $n_k v_k$ for all the planets in the system. For a given quantum-like number configuration $\{n_b,\, n_c,\, \cdots \}$, this sets the fundamental speed $v_F$ equal to the average of the $n_k\cdot v_k$ with $k\in\{b,c,d,\cdots\}.$

It is clear that several minima are encountered and that considering arbitrarily large values of $v_F$ would result in arbitrarily accurate alignments. Small values of $v_F$ result in different planets sharing the same quantum-like number. Inversely, large values of $v_F$ correspond to configurations in which the principal quantum-like numbers are not consecutive integers. We restrict the search for optimal fundamental speeds $v_F$ to small enough values to ensure that the smallest difference between successive quantum-like numbers is not more than two. 

The optimal fundamental speed and sequence of quantum-like numbers are then the ones resulting in the smallest sum of squared differences between $v_F$ and the closest orbital speed integer multiples $n_k\cdot v_k$ in the above described range of $v_F$ values as illustrated in Figure \ref{assignement} in the next Section. 

\subsection{Accidental probabilities }\label{accidentals}
With the principal quantum-like number assignment algorithm proceeding by optimization, it is important to evaluate the chance probability for a planetary system to accidentally fit the quantum-like prescription discussed above.

Consider a given inverse fundamental speed $\eta_F=1/v_F$ for a planetary system in which the  $k^{th}$ planet is characterized by $\eta_k$. The assigned quantum-like number then is $n_k=\lfloor{\eta_k/\eta_F}\rceil$ \footnote{$\lfloor{u}\rceil$ is the integer closest to $u$, whether smaller or larger than $u$.}. We can then define the residual $\rho_k=|\eta_k/\eta_F-n_k|$. Under the null hypothesis of an absence of structuring in terms of integers, $\rho_k$ would be uniformly distributed over $[0,\frac 1 2]$. However, $v_F$ is not given. Instead it is calculated as the mean of the $n_k\cdot v_k$ and this biases the residuals toward zero. This can be quantified using a simple Monte Carlo simulation. For systems with $N$ planets, we draw $N$ random numbers uniformly distributed between $-\frac 1 2$ and $+\frac 1 2$. The mean of these numbers is calculated and then subtracted from the random numbers. When the absolute value of the result is larger than $\frac 1 2$, it is replaced with its complement to unity so as to ensure all the values are in the interval $[0,\frac 1 2]$. These numbers are distributed like the residuals for $N$ planet systems under the null hypothesis with the biases resulting from the optimization procedure described in Subsection \ref{nassignalgo}. For a given actual residual $\rho_k$, the chance probability $p_k$ for the residual to be accidentally smaller under the null hypothesis can be estimated as the fraction of the Monte Carlo distribution below $\rho_k$.

Now, considering a system of $N$ planets, the product of the individual planets accidental probabilities $p= \prod_{k=1}^N p_k$ would then be the chance probability for the system to accidentally follow the relation $n_k=\eta_k/\eta_F$ more closely for all individual planet. However, a penalty must be paid for the optimization search that was performed.

The optimization proceeds continuously over a finite range of  fundamental speed values $v_F$, testing a finite set of quantum-like number configurations, which can be counted. Let $\mathcal N_{C\!f\!g}$ be the number of configurations explored. Then, following binomial statistics, the a priori accidental chance probability for one or more configurations to correspond to a better fit than the best one found is $P=1-\left(1-p\right)^{\mathcal N_{C\!f\!g}}$.

We will use this accidental probability to characterize the goodness of fit  for each planetary system. In this article, the reported accidental probabilities are all based on monte-carlo simulations of 100,000 systems for each number of planets. It should be noted that this probability does not depend on the quantum-like number sequence to be consecutive or to include multiplets of planets sharing the same assignment or not. 

\section{Solar system}\label{solar}
Table \ref{solarplanets} lists the names and orbital periods of the planets in the Solar system with the corresponding values of $\eta$, calculated with $M=1\,M_\odot$. In addition to the more familiar major planets, the most important minor planets are also included. Inspections of the differences between consecutive values of $\eta$ leads to the identification of two sections in the solar system. The first, the {\it inner solar system}, extends from Mercury to Mars, possibly including the minor planets Vesta, Ceres, Pallas and Hygiea.  The second, {\it the outer solar system}, extends from Jupiter to Neptune and possibly includes the minor planets Pluto, Haumea, and Makemake. We will consider each sub-system individually starting with the {\it inner solar system}. In this study, we do not include the trans-Neptunian system, which is composed of objects all considered as minor planets with eccentric orbits but they were also  shown to inscribe themselves in the resolution-scale relativistic macro-quantization picture of the Solar System \cite{nottale1993}.  

\begin{table}
\begin{center}
\begin{tabular}{ l c  c }
\hline \hline
Name.                & Period (yr)    & $\eta$ (yr/AU)     \\
\hline
Mercury (\Mercury)   & $0.240842$     &0.09902    \\
Venus (\Venus)       & $0.615186$     &0.13536    \\
Earth (\varTerra)    & $1.00$         &0.159155   \\
Mars (\Mars)         & $1.88079$      &0.196457   \\
Vesta                & $3.62994$      &0.244598   \\
Ceres                & $4.60397$      &0.264767  \\
Pallas               & $4.61046$      &0.264892  \\
Hygiea               & $5.56814$      &0.282092  \\
\hline
Jupiter (\Jupiter)   & $11.859 $      &0.362941  \\
Saturn (\Saturn)     & $29.4566$      &0.491527  \\
Uranus (\Uranus)     & $84.3219$      &0.697911  \\
Neptune (\Neptune)   & $164.788$      &0.872560  \\
Pluto                & $247.94$       &0.999852  \\
Haumea               & $283.12$       &1.045065  \\
Makemake             & $306.21$       &1.072737  \\
Eris                 & $558.046$      &1.310318   \\
\hline  
\end{tabular}
\caption{Table of the major planets and most important minor planets in the solar system with their orbital period and the corresponding $\eta$ values. The inner and outer sections of the solar system are separated by a horizontal line. \label{solarplanets}}
\end{center}
\end{table}

\subsection{Inner solar system}

Focusing on the major planets of the inner solar system in Table \ref{solarplanets}, we see that the values of $\eta$ are close to being equally spaced with the mean difference between successive $\eta$ values being $\langle{\delta\eta}\rangle=\frac{\eta_\text{\Mars}-\eta_\text{\Mercury}}{3}\approx0.0325\,\rm yr/AU$. With $\eta_\text{\Mercury}\approx0.09902\,\rm yr/AU$, this suggest the {\it quantum-like numbers} assignments $n_\text{\Mercury}=3$, $n_\text{\Venus}=4$, $n_\text{\varTerra}=5$, and , $n_\text{\Mars}=6$. 

\begin{figure}[!ht]
\begin{center}
\includegraphics[width=8cm]{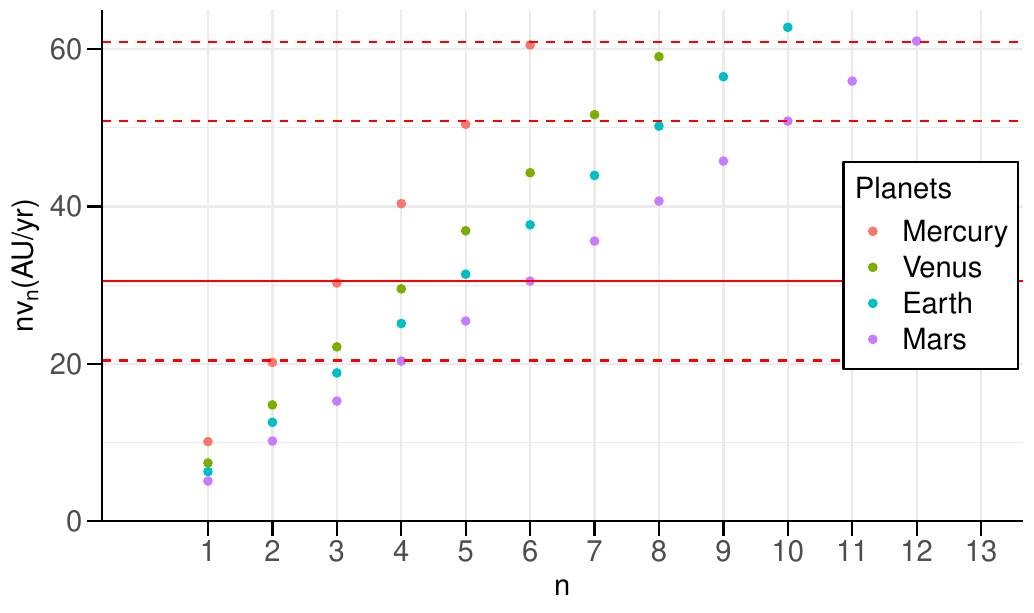}
\includegraphics[width=8cm]{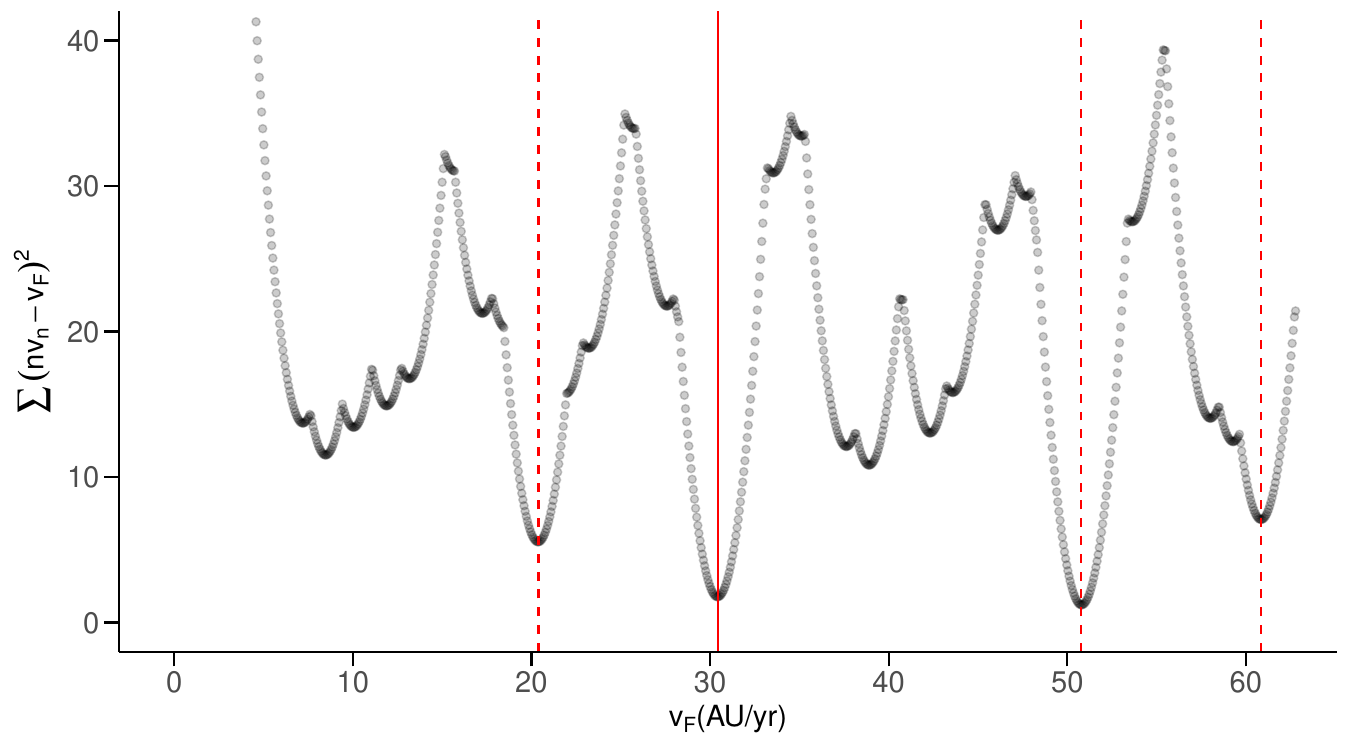}
\end{center}
\caption{Illustration of the quantum-like number assignment algorithm applied to the four planets of the inner solar system (Mercury, Venus, Earth, and Mars). In the upper panel, each series of colored dots shows the respective mean orbital speeds multiplied by the integers on the horizontal axis. The lower panel shows the sum of squared difference between a tentative fundamental orbital speed and the closest integer multiple of the orbital speed. The four dashed vertical lines correspond to the four deepest local minima also indicated in the upper panel to identify optimal quantum number assignments. One of the two lowest minima corresponds to $v_F=30.47\,\rm AU/yr$ and the quantum-like numbers 3, 4, 5 and 6 assigned to Mercury, Venus, Earth and Mars respectively. The maximal value of the tested fundamental speed was based on the observation that for $v_F= 60.88\,\rm  AU/yr$ already, principal quantum-like numbers do not have consecutive integer values. }
\label{assignement}
\end{figure}

This should be compared to what is obtained when applying the algorithm described in Section \ref{nassignalgo} as illustrated in Figure \ref{assignement}. The sum of the squared difference between integer multiples of the orbital speeds  and a fundamental speed reaches four main minima over the considered interval. Starting from the smallest values of $v_F$, the first minimum, $v_F= 20.42\,\rm AU/yr$, with a value of $5.56\rm\,(AU/yr)^2$, corresponds to Venus and Earth sharing the same quantum-like number 3 while Mercury and Mars are assigned quantum-like numbers 2 and 4 respectively. The last minimum, $v_F=60.88\,\rm AU/yr$, with a value of $7.09 \rm\, (AU/yr)^2$ is associated with the sequence 6, 8, 10, and 12 for Mercury, Venus, Earth, and Mars respectively. 

\begin{figure}[!ht]
\begin{center}
\includegraphics[width=8cm]{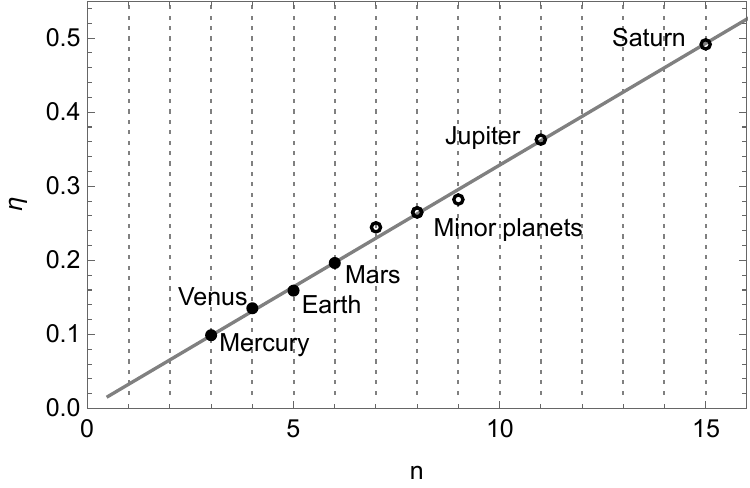}
\end{center}
\caption{$\eta$ is shown as a function of the assigned {\it quantum-like number} $n$ for the inner solar system. Solid dots represents major planets used in the optimization. Open circles represent minor planets and major planets that are not used in the optimization. The line represents a proportionality relation between $n$ and $\eta$ with a equal to the optimal inverse orbital speed $\eta_F=0.03286\,\rm yr/AU$.}\label{innersolar}
\end{figure}

The two other minima are quite deeper with values of  $1.78  \rm\, (AU/yr)^2$ and $1.22 \rm \,(AU/yr)^2$ for $v_F=30.47\,\rm AU/yr$ and $v_F=50.83\,\rm AU/yr$ respectively with the principal quantum-like number sequences 3, 4, 5, 6 and 5, 7, 8, 10. The first of these two minima corresponds to the inner-solar system description obtained at the beginning of this section. The accidental probability is $P=7.2\times 10^{-3}$. Up to a factor two on $v_F$ and the assigned principal quantum-like numbers, it also corresponds to the minimum with $v_F=60.88\,\rm AU/yr$. It is the description presented in Figure \ref{innersolar}, which distinguishes itself from the other minimum by resulting in the assignment of consecutive integers to all the planets used in the optimization: $n=3$ for Mercury, $n=4$ for Venus, $n=5$ for Earth, and $n=6$ for Mars. The points nicely follow the relation $\eta=\eta_F\cdot n$ with $\eta_F=(0.03286\pm0.0007)\,\rm yr/AU$. The solar system orbital periods are all known with great precision so the errors reported here for values of $\eta_F$ are obtained as the standard deviations of the ratios $\eta/n$ about the mean. This corresponds to $v_F=\frac{1}{\eta_F}=(30.43\pm0.66)\,\rm AU/yr$ or $v_F=(144\pm3)\times10^3\,\rm m/s$ as already reported by Nottale \cite{nottale2000,nottale2011}. Using Equation \ref{bohrv} with $n=1$, we have $\mathcal D=\frac 1 2 \eta_F G M$, which gives $\mathcal D=(0.648\pm0.014)\,\rm AU^2/yr$. Using the same value of $v_F$, we can continue with consecutive integers for Vesta with $n=7$ (0.45)\footnote{Here, numbers between parentheses are the respective residuals.}, both Ceres and Pallas with $n=8$ (0.068 and 0.072), and Hygiea with $n=9$ (0.40). For Jupiter and Saturn,  $n_\text{\Jupiter}=11$ (0.060) and $n_\text{\Saturn}=15$ (0.021). Multiplying the accidental probability for the four major planets by twice the respective residuals of these additional objects not used in the optimization gives an accidental probability of $5.4\times 10^{-7}$. It is remarkable that objects,  Ceres, Pallas, Jupiter, and Saturn, line up so well while they were not considered in the optimization. It is also remarkable that this is obtained with consecutive integers for the first eight objects. Vesta and Hygiea seem to be further away from the linear relation. This could be related to the transition between the inner and outer solar systems. It should also be noted that {\it orbital} $n=2$ is not occupied. From Equation \ref{bohrt}, $n=2$  would correspond to an orbital period $T_2\approx \left(\frac{2}{3}\right)^3 T_\text{\Mercury}\approx 26\,\rm days$.  The fact that $n=1$ is not occupied either could be expected if one thinks about the fact that the hydrogenoid orbital $n=1$ is not associated with any orbital angular momentum, a situation not compatible with the non-degenerate closed orbit of a classical material object.  
 
Although the above description is in very good compliance with the expectations from Resolution-Scale Relativity, it does not correspond to the strict minimum of the sum of the squared difference between integer multiples of the orbital speeds and a fundamental speed, obtained for $v_F=50.83\,\rm AU/yr$ with quantum-like numbers 5, 7, 8, and 10 assigned to planets Mercury through Mars respectively for a slightly degraded accidental probability $P=8.4\times 10^{-3}$. This illustrates the fact that the algorithm may converge on a slightly deeper minimum providing a less compelling model. We accept this and in the rest of the article, the algorithm will be applied blindly without any further scrutinization in search for more compelling models.

\subsection{Outer solar system}
\begin{figure}[!ht]
\begin{center}
\includegraphics[width=8cm]{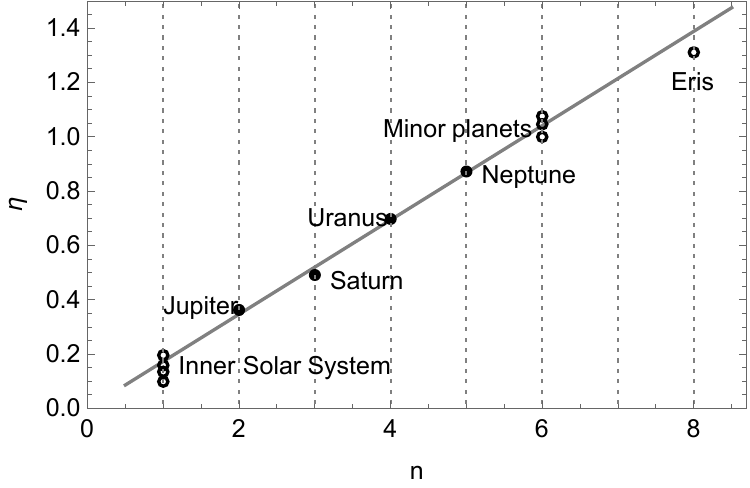}
\end{center}
\caption{Same as Figure \ref{innersolar} but for the outer solar system. Solid dots are for the four major planets while open dots are for the four major planets of the inner solar system and for the minor planets of the outer solar system. The line represents a proportionality relation between $n$ and $\eta$ with a slope corresponding to the inverse fundamental speed $\eta_F = (0.1736\pm0.0063)\,\rm yr/AU$}\label{outersolar}
\end{figure}

The sum of the squared difference between integer multiples of the orbital speeds and a fundamental speed for the major objects of the outer solar system (Jupiter, Saturn, Uranus, and Neptune) was minimized, resulting unambiguously in a fundamental speed  $v_F=5.77\,\rm AU/yr$, corresponding to consecutive principal quantum-like numbers from $n=2$ to $n=5$, respectively for Jupiter, Saturn, Uranus, and Neptune as shown on Figure \ref{outersolar}, with an accidental probability of $P=3.0 \times 10^{-2}$. The corresponding outer solar system fundamental speed is $v_F=(5.76\pm0.21)\,\rm AU/yr$, or $v_F=(27.3\pm1.0)\times 10^3\,\rm m/s$.  
It is $5.28\pm0.22$ times smaller than the fundamental speed found for the inner-solar system, and corresponds to an effective diffusion constant $\mathcal D=(3.43\pm0.12)\,\rm AU^2/yr$.

Using this fundamental orbital speed, the inner-solar system collectively corresponds to $n=1$, and the minor-planets Pluto, Haumea, and Makemake are all well accommodated by $n=6$. However, for Eris, the choices $n=7$ and $n=8$ are equally bad.   This is reminiscent of the situation of minor-planets at the interface between the inner and outer-solar systems with Eris being on the outskirt of the outer-solar system and on the inner-edge of the trans-neptunian solar system. Additionally, this time again, it is remarkable that, although they were not part of the optimization, the major-planets of the inner-solar system and minor-planets of the outer-solar system, with the exception of Eris as already discussed, all fit the model with consecutive integers.  

\section {Extra-solar systems}\label{fiveandmoreplntssyst}

The NASA Exoplanet Archive \cite{han2014} provides orbital the parameters for 4,336  planetary systems, one with eight characterized planets, one with seven, 11 with six, 27 with five, 75 with four, 207 with three, 648 with two, 3364 with one. 
We decided to focus on the 115 systems with four or more planets, to which we systematically applied the analysis described in Section \ref{analysis} and applied to the solar systems in Section \ref{solar}.  

\begin{figure}[!ht]
\begin{center}
\includegraphics[width=8cm]{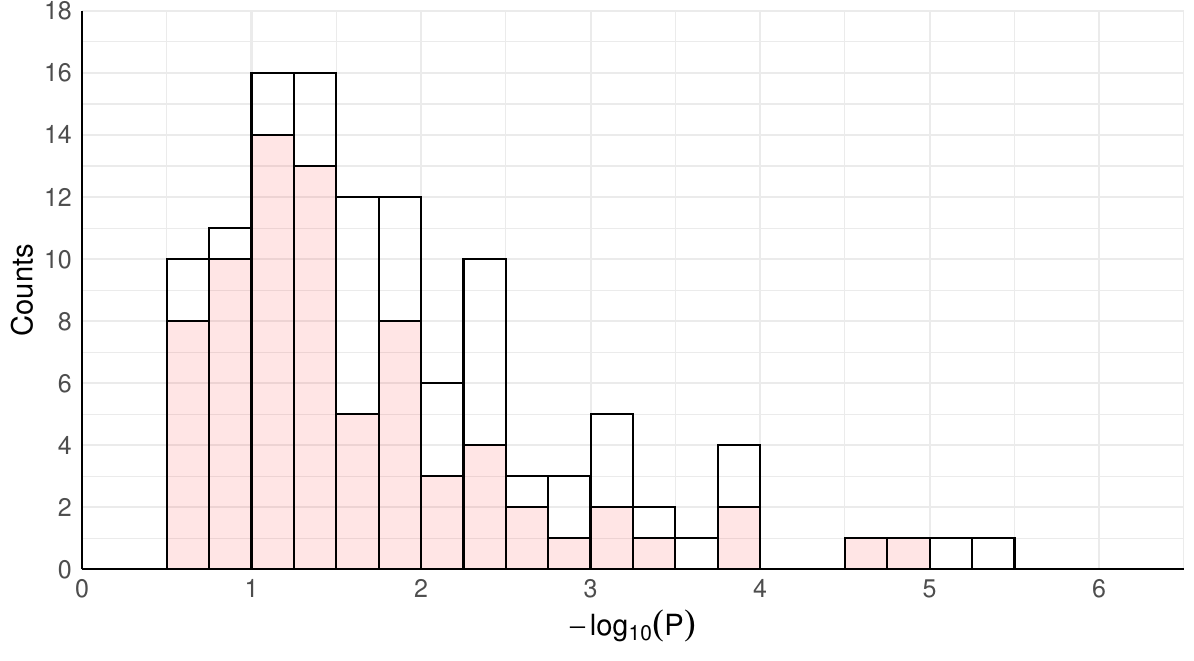}
\end{center}
\caption{Histogram of $-\log_{10}(P)$, where $P$ is the accidental probability described in Sub-section \ref{accidentals}, for the 115 systems with at least 4 planets (clear) and for the 75 systems with just four planets (shaded). Of the systems with 4 planets, 58 have an accidental probability of more than $10^{-2}$ while of the 40 systems with 5 or more planets, only 19 (47\%) have an accidental probability of more than $10^{-2}$. To characterize the propensity of our 115 planetary systems to comply with the scale relativistic macro-quantization, we note that 38 systems (33\%) have an accidental probability of less that $10^{-2}$,  16 (14\%) less than $10^{-3}$, 4 (3.5\%) less than $10^{-4}$, and 2 (1.7\%) less than $10^{-5}$. 
}
\label{probadistri}
\end{figure}

Figure \ref{probadistri} shows the distribution of accidental probabilities for the optimal configurations for systems with more than 4 planets and distinguishes systems with only 4 planets. Of the 75 systems with 4 planets, 58 (77\%) have an accidental probability of more than $10^{-2}$ while of the 40 systems with 5 or more planets, only 19 have an accidental probability of more than $10^{-2}$. This justifies our choice to restrict ourselves to systems with no less than 4 planets for this analysis. Turning to the characterization of the propensity of the 115 planetary systems to comply with the scale-relativistic macro-quantization, we note that 38 systems (33\%) have an accidental probability of less that $10^{-2}$,  16 (14\%) less than $10^{-3}$, 4 (3.5\%) less than $10^{-4}$, and 2 (1.7\%) less than $10^{-5}$. This corresponds to a clear excess of low accidental probability values. 

It should be noted that the optimization procedure and the accidental probability are insensitive to the consecutive character of the set of quantum-like numbers in a system. It is then remarkable that, of the 401 pairs of successive quantum-like numbers within individual planetary systems, 241 (60\%) are different by one and 89 (22\%) are different by two. That is, 82\% of the pairs of successive quantum-like numbers are different by no more than two, while the $v_F$ optimization extends up to the value for which the system's smallest difference between consecutive quantum-like numbers is equal to two. 

In the same thread, we notice that, of the 115 planetary systems, altogether 34 (29\%) are described by a series of consecutive integers. Distinguishing systems with different numbers of planets, this corresponds to 25 (33\%) of 75 systems with 4 planets, 7 ( 26\%) of 27 systems with 5 planets, and 2 (18\%) of 11 systems with 6 planets. Altogether, there are 101 systems (88\%) in which at least half of the quantum-like numbers are part of consecutive sequences. 
 
Only one system is not assigned any consecutive quantum-like numbers. It is Kepler-132 for which the two first planets are assigned a common quantum-like number, and the two others are separated by gaps greater than one (the sequence being 5, 5 ,7, and 13 with an accidental probability of $P=4.1 \times 10^{-2}$).

\begin{figure}[!ht]
\begin{center}
\includegraphics[scale=0.47]{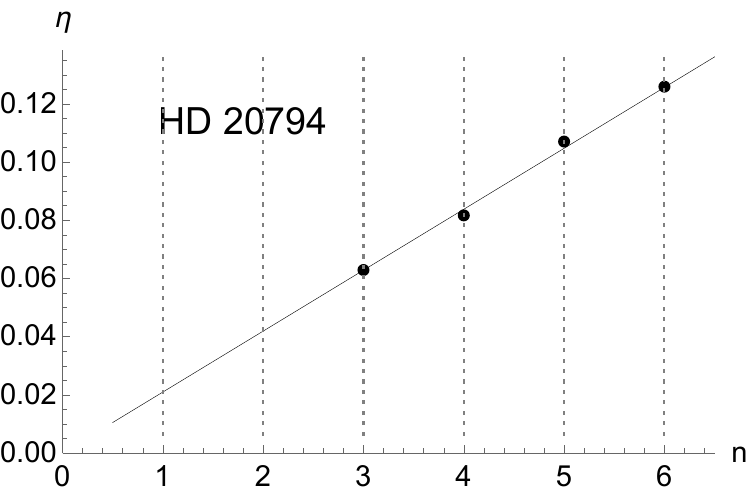}
\includegraphics[scale=0.47]{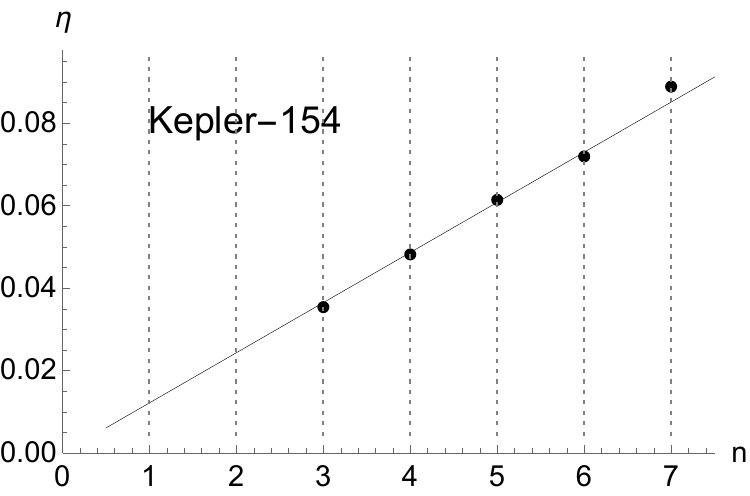}
\includegraphics[scale=0.47]{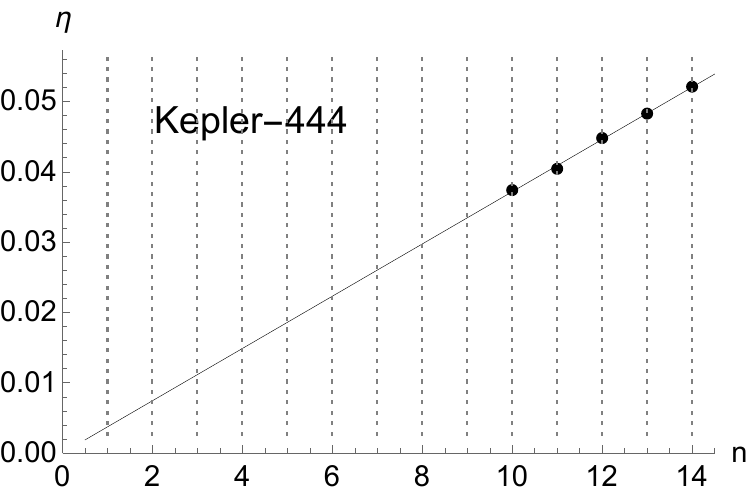}
\includegraphics[scale=0.47]{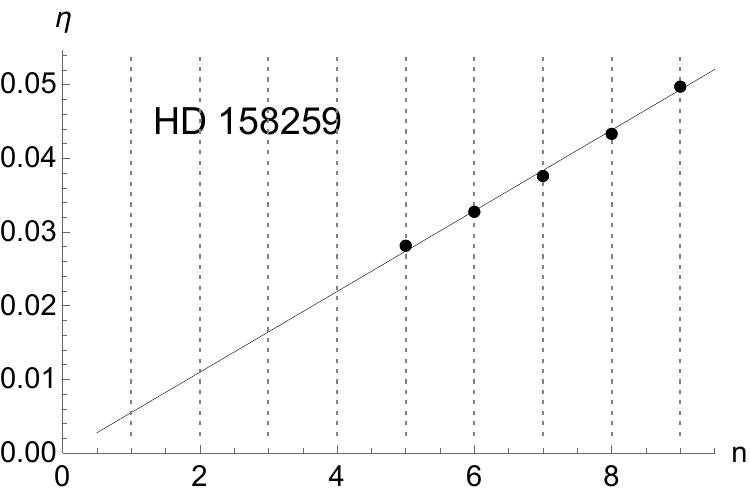}
\end{center}
\caption{Principal quantum-like number assignments for planetary systems HD 20794, Kepler-154, Kepler-444, and HD 158259, which are all described by consecutive values of principal quantum-like numbers, with respective accidental probabilities of $4.7\times 10^{-4}$, $1.5 \times 10^{-2}$, $1.0\times 10^{-3}$, $1.7 \times 10^{-2}$.}\label{consecexamples}
\end{figure}

Four examples of planetary systems for which the optimal quantum-like numbers form consecutive sequences are shown in Figure \ref{consecexamples}. Among all systems with a sequence of consecutive quantum-like numbers, only one extends beyond $n=10$. It is the Kepler-444 five planet system, with the sequence 10, 11, 12, 13, and 14 and an accidental probability $P=1.0\times 10^{-3}$. 

\begin{figure}[!ht]
\begin{center}
\includegraphics[scale=0.45]{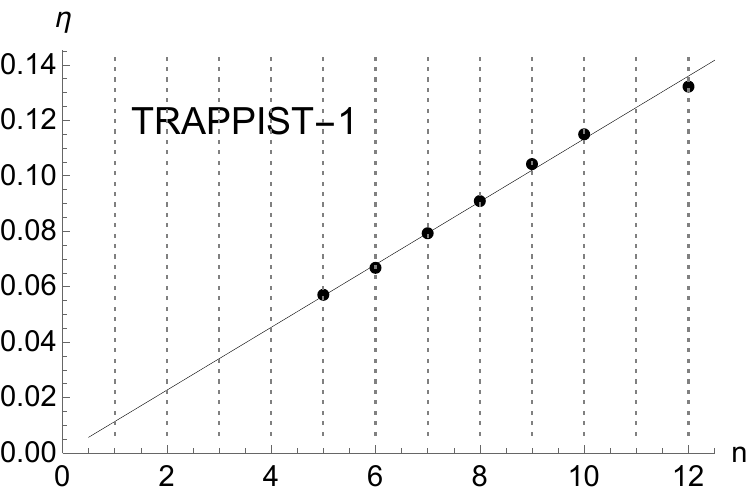}
\includegraphics[scale=0.45]{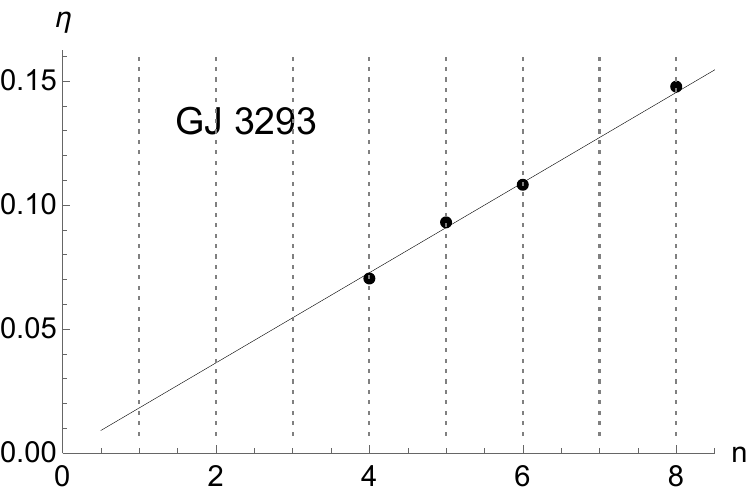}
\includegraphics[scale=0.45]{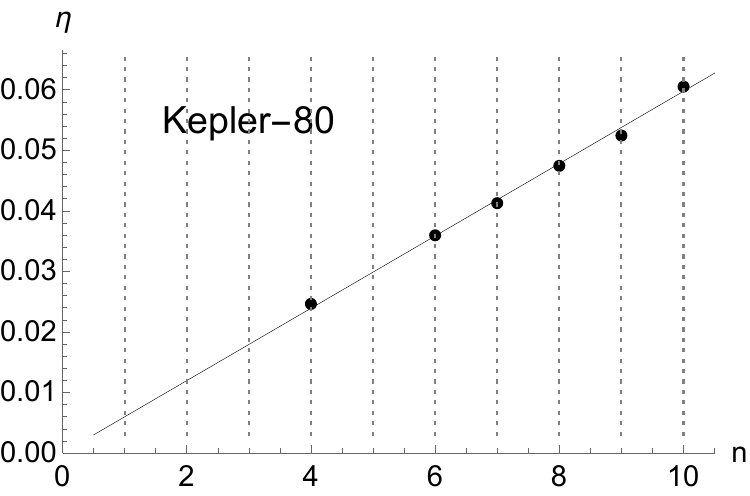}
\includegraphics[scale=0.45]{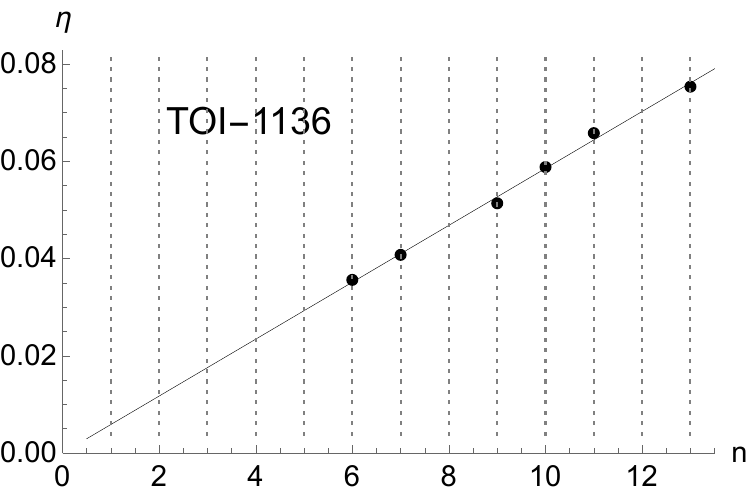}
\end{center}
\caption{Principal quantum-like number assignments for planetary systems TRAPPIST-1, GJ\,3293, Kepler-80, and TOI-1136, which are all described by consecutive values of principal quantum-like numbers with one or two gaps and the following respective accidental probabilities  $1.0 \times 10^{-4}$, $0.16$, $5.2 \times 10^{-3}$, and $1.7 \times 10^{-2}$}.\label{holeysyst}
\end{figure}

Similarly, Figure \ref{holeysyst} presents four examples with sequences of consecutive numbers with one or two gaps. The gaps could correspond to planets that have yet to be detected.  

\begin{figure}[!ht]
\begin{center}
\includegraphics[scale=0.4]{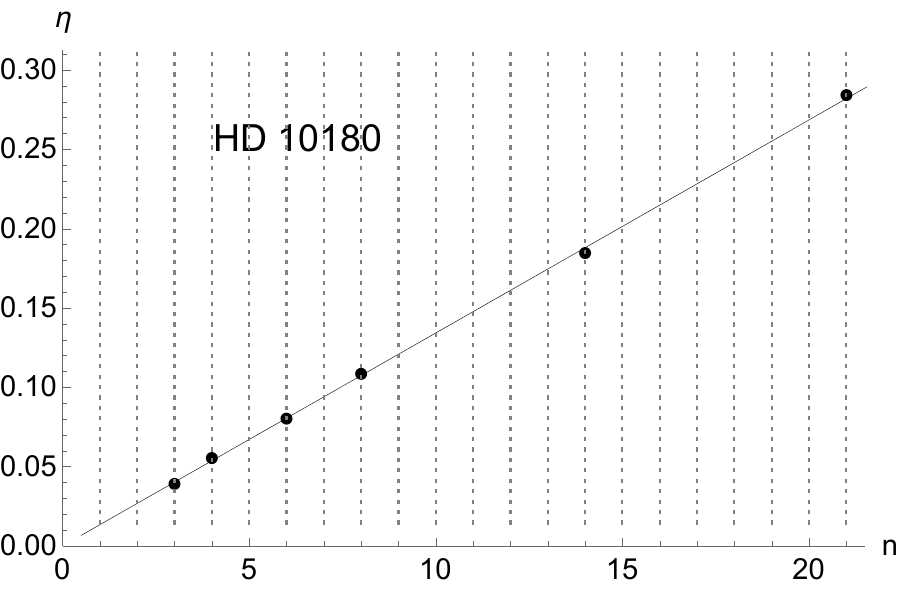}
\end{center}
\caption{Principal quantum-like number assignments for the HD10180 system, which could have a hierarchical structure similar to what was discussed in the case of the solar System. The accidental probability is $P= 7.5 \times 10^{-3}$.  }\label{inoutexamples}
\end{figure}

Most assigned quantum-like numbers are inferior to 16 (93\%). The few larger quantum-like numbers occur in systems that could come under a hierarchical description similar to what we have seen in the solar system. One beautiful example of this is the HD 10180 system with quantum-like numbers 3, 4, 6, 8, 14 and 21 as shown in Figure \ref{inoutexamples}, with an accidental probability of $P= 7.5 \times 10^{-3}$ and a fundamental speed $v_F= 74.5\rm\,AU/yr$. The similarity with the solar system is clear. The first four planets could constitute an inner-system and the last two an outer-system. In fact, focusing on the outer-system leads to an alternative picture with the first four planets sharing $n=1$, the fifth in $n=2$ and the last in $n=3$ with respective residuals of $0.026$ and $0.039$, for a fundamental speed $v_F=10.7\rm\,AU/yr$ (value based on the two outer planets exclusively), which is $\sim7$ times smaller than for the inner system.

\begin{figure}[!ht]
\begin{center}
\includegraphics[scale=0.6]{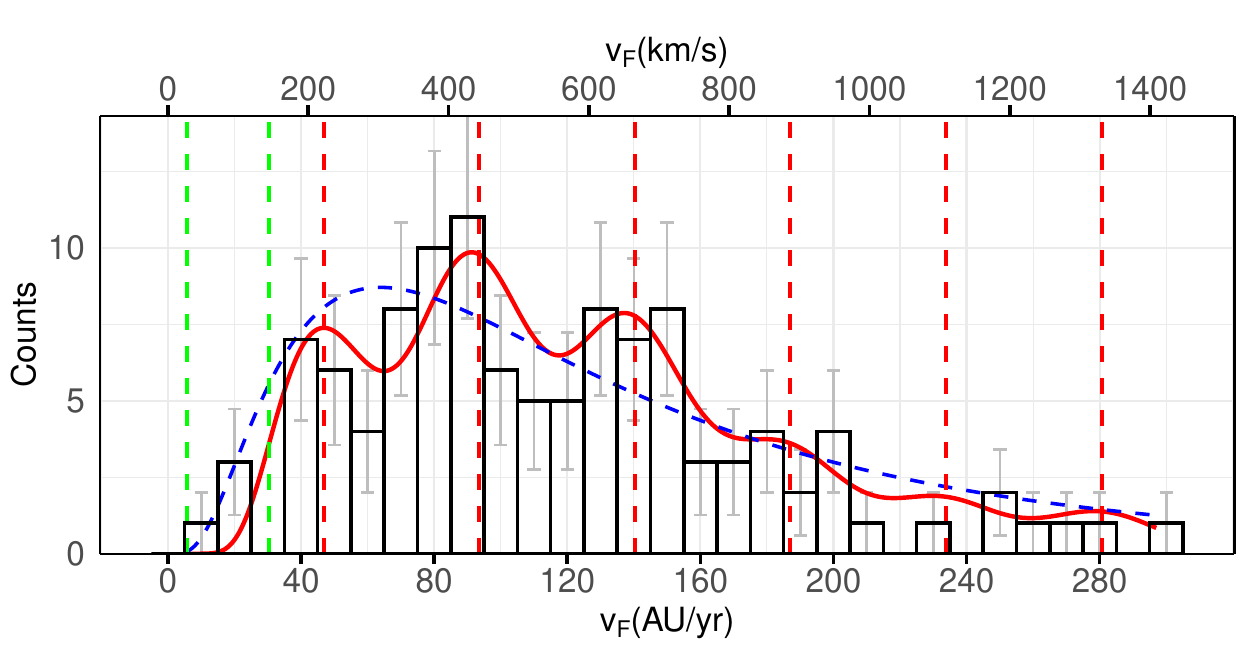}
\end{center}
\caption{Distribution of the extra-solar planetary systems fundamental speeds. The distribution is presented with error bars obtained by considering the bin counts as  Poisson random variables. A fit (the solid red curve) of a sum of log-normal distributions with modes $k v_0$ and log of speed scales $\sigma/k$ with $k$ an integer has been performed, yielding the measurements $v_0=(46.8\pm1.1)\,\rm AU/yr$, and $\sigma=(0.36\pm 0.044)$. The dashed vertical red lines show the values of $kv_0$. The fit with a single log-normal function  performed for comparison is shown by the dashed blue curve (see text for details). With $v_F=\frac{1}{\eta_F}=(30.43 \pm0.66)\,\rm AU/yr$ (rightmost vertical dashed green line), the inner solar system falls within the first peak. However, the outer solar system, with $v_F=(5.76\pm0.21)\,\rm AU/yr$ (leftmost vertical dashed green line) does not correspond to any peak, suggesting that, if the fundamental speed distribution has some universal character, it would be structured in a way more complex than the simple periodicity proposed here. 
}\label{fundveldistri}
\end{figure}

With the above results suggesting the compliance of planetary systems with the scale relativistic quantum-like dynamics, we can now turn to the values of the fundamental speeds whose distribution is presented in the histogram of Figure \ref{fundveldistri} for the 115 systems. Two fundamental speeds are out of range.  They are obtained for the systems DMPP-1 and Kepler-1542,  respectively with $434\,\rm  AU/yr$ and $585\,\rm AU/yr$.

The distribution fits a log-normal law $P(v_F;v_0,\sigma)$\footnote{$P(v_F;v_0,\sigma) = (v_F\sigma\sqrt{2\pi})^{-1}\exp(-(\ln(v_F/v_0)-\sigma^2)^2/2\sigma^2)$} (dashed curve in Figure \ref{fundveldistri}) with mode $v_0=64\,\rm AU/yr$ and log of speed scale $\sigma=0.78$. The fit is acceptable with a reduced $\chi^2$ of 0.63. However, we notice that the distribution deviates from the log-normal law in a way that could be periodic. So we fit a sum of equally spaced log-normal functions of the form $\sum_{k=1}^{6} a_k  P(v_F; kv_0, \sigma/k)$ where $kv_0$ is the mode of the $k^{th}$ component and $\sigma/k$ its log of speed scale. With a reduced $\chi^2$ of 0.52, the fit provides the measurement $v_0=(46.8\pm1.1)\,\rm AU/yr$ or $v_0=(218.0\pm4.7)\,\rm km/s$, with $\sigma=(0.36\pm 0.044)$. The fit has been performed without the two leftmost bins close to $v_F=0$ in Figure \ref{fundveldistri} which are not well accommodated by a log-normal profile. With $v_0\sigma = 16.9 \,\rm AU/yr$, the peaks are well separated (solid curve in Figure \ref{fundveldistri}) with respective amplitudes  greater than five standard deviations for the first peak ($k=1$), six for the second and third ($k=2$ and $k=3$), and four for the fourth ($k=4$), while the last two ($k=5$ and $k=6$) are not statistically significant.   

The possible appearance of a universal character in the distribution of fundamental speeds was not expected and provides an additional indication that Resolution-Scale Relativity is at play in planetary systems. Nevertheless, in the next Section, we return to considering systems individually to test if mutual orbital resonance could be responsible for the quantum-like structuring we observe.  

\section{Orbital resonances}\label{resonance}

Planetary systems being composed of several mutually interacting objects with different periods, their stability depends on resonance effects in complicated ways. When two orbital periods are in a ratio close to a simple rational, the relative positions of the two objects after each orbit completion by the inner one are periodic and the effect of the mutual interaction builds up instead of averaging out \cite{peale76,agol2005}. Inversely, some stable mutual locking can happen between eccentric orbits with orbital periods whose ratios are very close to simple rational numbers. 

\begin{figure}[!ht]
\begin{center}
\includegraphics[scale=0.4]{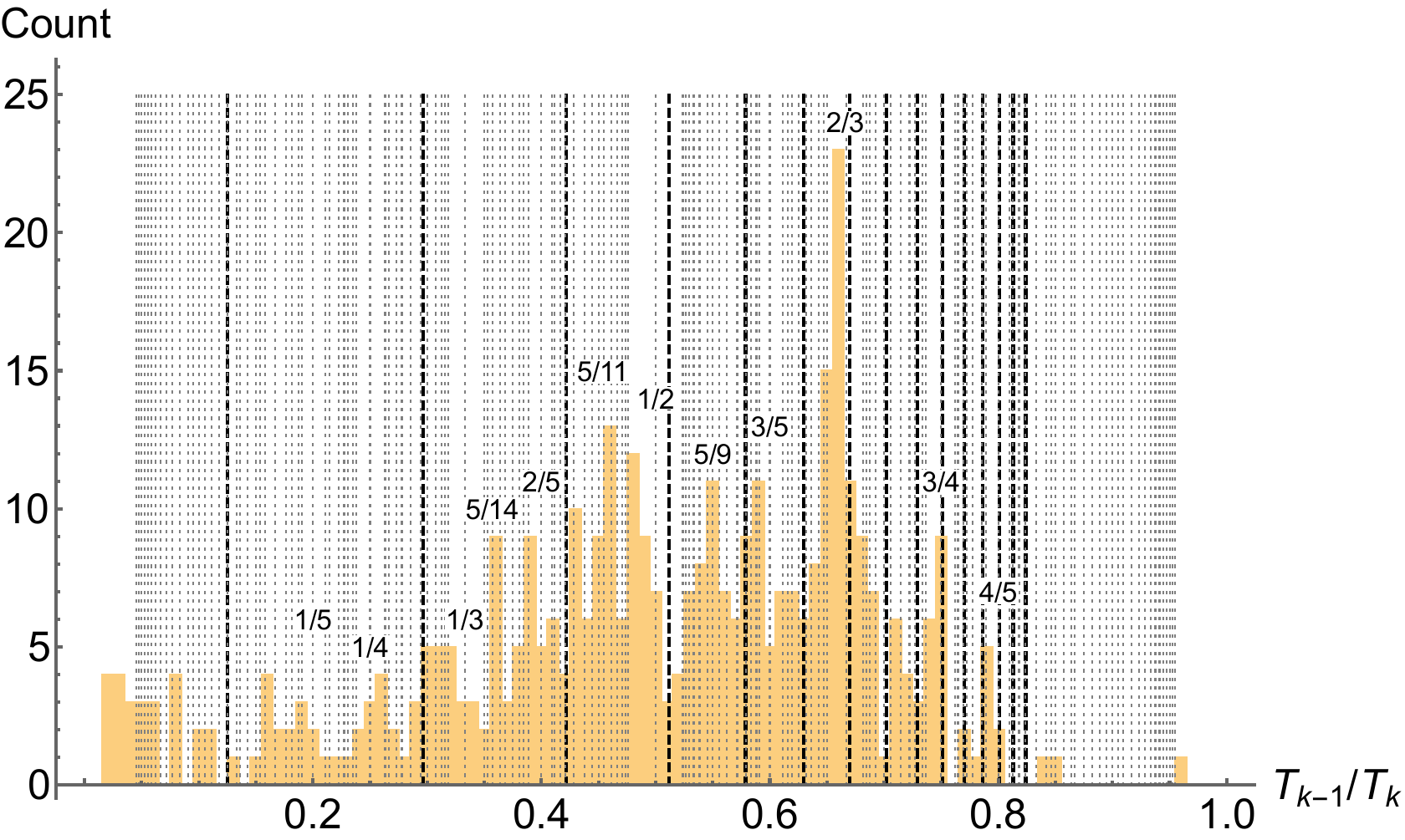}
\end{center}
\caption{The histogram of the orbital ratios between consecutive planets is compared to the simple rationals $p/q$ with $q>p$ and $q<23$ marked by the vertical dotted lines. The vertical dashed line indicate values of $p/q$ for which resonance conditions would line up with the macro-quantization description. See text for details.  }\label{resonance_histo}
\end{figure}

The manifestation of these effects can be seen in Figure \ref{resonance_histo} showing the distribution of the ratio between consecutive orbital periods in systems with at least four identified planets. The histogram is compared to simple rationals $p/q$ with $q>p$ and $q<23$ represented by vertical dotted lines. It is clear that peaks in the orbital period ratios distribution tend to occupy the gaps surrounding the simplest rational numbers such as $2/3$, $1/2$, or $3/4$.  

If a planetary system included a succession of planets with consecutive orbital period ratios equal to each others, the choice of {\it quantum-like principal numbers} could just amount to lining up with a geometric progression $T_{k-1}/T_{k}=p/q$ of the orbital periods $T_k$ with $p/q$ representing a rational number with $q>p$. This would result in $\eta_{k-1}/\eta_k=\left(p/q\right)^{1/3}$, but for a confusion to really emerge it would also require that $\eta_{k}-\eta_{k-1}\approx \eta_k/k$, which would imply  $p/q=\left({\frac{k-1}{k}}\right)^3$  for $k\geq 2$. These values of $p/q$ are indicated by the vertical thick dashed lines on Figure \ref{resonance_histo}. It appears that $p/q$ for $k=8$ and $k=11$ quite precisely line up with peaks for resonance factors $2/3$ and $3/4$ respectively. 

While these peaks deserve attention, in fact, we do not have many systems with sequences of identical resonance factors. Considering the rational numbers $p/q$ with $q\le10$ that are the closest to orbital period ratios between consecutive planets, of our 115 systems, we find one with the same factor repeated three times, two with the same factor repeated twice and sixteen with the same factor repeated once. So resonances cannot generally be responsible for a fake signature of scale-relativistic quantization. Systems with resonances are nevertheless interesting to look at. 

\begin{figure}[!ht]
\begin{center}
\includegraphics[scale=0.45]{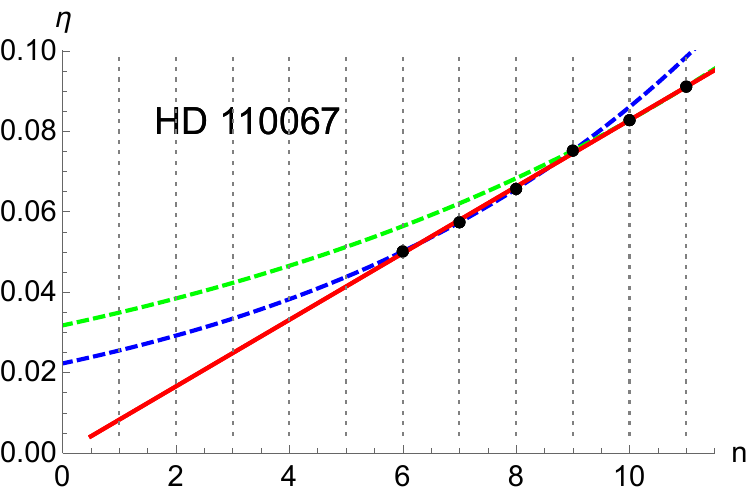}
\includegraphics[scale=0.45]{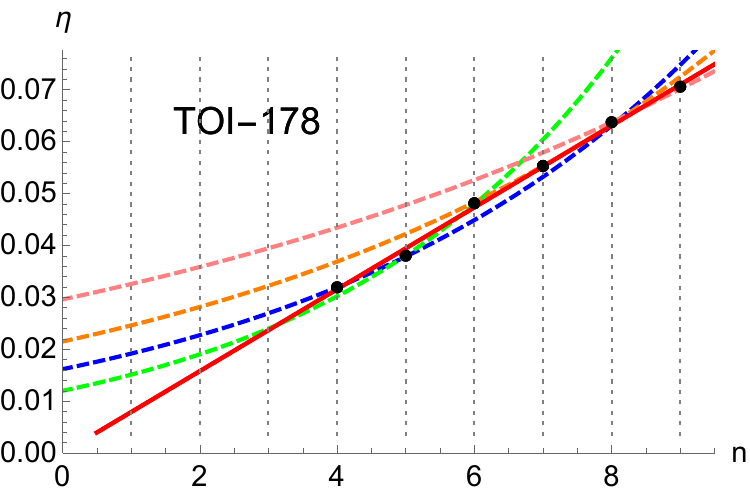}
\end{center}
\caption{For systems HD\,110067 (left) and TOI-178 (right), the values of $\eta$ for the different planets is shown as a function of the assigned principal quantum-like numbers. The solid red line indicates the proportionality relation with slope corresponding to the inverse fundamental speed $\eta_F$ in each case. The dashed lines represent the exponential relation based on the orbital resonance ratios for the recognized resonances. For HD\,110067, the blue and green curves represent the exponential of bases 3/2 and 4/3 respectively. For TOI-178, the blue, green, orange, and pink curves represent exponentials of bases 5/3, 2/1, 3/2 and 4/3 respectively.}\label{resonance_exemples}
\end{figure}

System HD\,110067 attracted a lot of attention for its highly resonating configuration. Its orbital period ratios between successive planets are as follows: 
$T_c/T_b=1.500349$, 
$T_d/T_c=1.500664$, 
$T_e/T_d=1.500666$, 
$T_f/T_e=1.333369$, and 
$T_g/T_f=1.333947$, precisely corresponding to 3/2 three times in a row and 4/3 two times. The $\eta$ versus $n$ graph for HD\,110067 is presented on the left panel of Figure \ref{resonance_exemples}, with the exponential laws connecting the first four and the last three planets respectively.

Similarly TOI-178 seems to include resonances with less repetitions of the same orbital period ratios, which are also less close to simple rational numbers: 
$Tc/Tb=1.692$ ( 5 : 3 ),
$Td/Tc=2.025$ ( 2 : 1 ),
$Te/Td=1.52$ ( 3 : 2 ),
$Tf/Te=1.53$  ( 3 : 2 ), and
$Tg/Tf=1.36$ ( 4 : 3 ). The $\eta$ versus $n$ graph for TOI-1136 is presented on the right panel of Figure \ref{resonance_exemples}, with the exponential laws connecting each pair of planets by the exponential law corresponding to their orbital period ratio. 

While systems HD\,110067 and TOI-178 include orbital resonances, both are well described by the relation $\eta_n=n\,\eta_F$ with all consecutive quantum-like principal quantum numbers for accidental probabilities of $4.4\times10^{-6}$ and $7.3 \times 10^{-4}$ respectively. Figure \ref{resonance_exemples} shows that this requires different orbital period ratios to be combined together so as to have the system comply with the scale relativistic macro-quantization . 

This is suggestive that, if quantum-like Resolution-Scale Relativistic structuring is indeed at play during planetary formation, and/or also during the long time-scale chaotic evolution of the system once the planets have formed and fell into a classical orbital description, resonance locking establishes itself with orbital period ratios preferentially close to those corresponding to the macro-quantum structure. At the same time, these mutual interaction effects operating over shorter time scales must cause some smearing of the macro-quantum structures and it is then remarkable that is seems we still get to observe them.

\section{Summary and conclusions}\label{closing}
We have used data from the NASA Exoplanet Archive to search for signatures of a hypothetical scale-relativistic macro-quantization in Keplerian systems. Namely, we used stellar masses and planetary orbital periods for 115 planetary systems with at least four planets. We calculated the inverse mean orbital speeds as the cube root of orbital periods divided by the stellar mass. In the Resolution-Scale Relativity framework, this quantity is expected to be an integer multiple of an inverse fundamental speed characteristic of each system (see Subsection \ref{bohr}). 

The analysis then proceeds by an optimization to assign a principal quantum-like number to each planet (see Subsection \ref{nassignalgo}) and to calculate an accidental probability for the data to fit the model with greater accuracy in the absence of macro-quantum structuring (see Subsection \ref{accidentals}). 

The analysis was first applied to the solar system, which fits a hierarchical structure distinguishing the inner (from Mercury through Mars) from the outer solar system (from Jupiter to Neptune) with respective fundamental speeds of $(144\pm3)\,\rm km/s$ and $27.3\pm1.0)\,\rm km/s$ for accidental probabilities of $7.2\times 10^{-3}$ and $3.0\times10^{-2}$.

The same analysis was then applied to NASA Exoplanet Archive data \cite{han2014} for the 115 systems with four or more planets (see Section \ref{fiveandmoreplntssyst}). We found a significant excess in the number of systems fitting the Resolution-Scale Relativistic macro-quantization for Keplerian systems with low accidental probabilities: 33\% of the systems have an accidental probability of less that $10^{-2}$,  14\% less than $10^{-3}$, 3.5\% less than $10^{-4}$, and 1.7\% less than $10^{-5}$. Additionally, there is an important proportion of consecutive quantum-like numbers. For example, we noticed that, independently from the accidental probabilities, 29\% of the systems are described by a series of consecutive integers and 88\% have at least half of the quantum-like numbers parts of consecutive sequences. We verified that only very few systems include repeating orbital period ratios that could be responsible for these results (See Section \ref{resonance}). In fact, we found examples of systems including resonances with orbital period ratios close to simple rational numbers changing from planet to planet while following the macro-quantization law. 

The fundamental speed of each system is the orbital speed corresponding to quantum-like number $n=1$. The fundamental speeds are distributed over a relatively broad range extending from less than $20\,\rm AU/yr$ ($100\,\rm km/s$) to more than $240\,\rm AU/yr$ ($1,200\,\rm km/s$) and can be described in terms of peaks centered on integer multiples of what would be a super-fundamental speed $v_0=(218.0\pm4.7)\,\rm km/s$ (see Figure \ref{fundveldistri}). Such a universal character to fundamental orbital speeds is striking. In the usual hydrogen atom described with the Bohr model, the speed of the electron in the orbital $n=1$ is the speed of light multiplied by the fine structure constant $\frac c {137}=2,188\,\rm km/s$, which is just ten times higher than the super fundamental speed we identify. Inversely, we can identify the planetary system quantum-like gravitational fine structure constant as $v_0=\frac c {1,376\pm 30}$. We do not have a theoretical framework to account for a universal super-fundamental speed or even to express it in terms of the speed of light. In fact if there is a universal character to the distribution of fundamental speeds, it is certainly more complicated than the simple periodicity in terms of a super-fundamental speed as suggested here. This is already suggested by hierarchical systems such as the solar system or HD10180, in which the outer system fundamental speed is too small to simply fit in the description in terms of integer multiples of the super-fundamental speed. There is a clear observational bias in favor of planets with high orbital speeds. Planets with orbital periods and speeds comparable to those in the outer-solar system are not likely to be detected. So, with this data, we are observing the tail of the orbital speed distribution, which is limited by the physical extent of the system's central stars. It is however interesting to note that while from Bohr's radius to the Astronomical Unit there are twenty-two orders of magnitude, from atomic speeds to planetary fundamental speeds there is just one order of magnitude. Noting that galaxy rotation curves plateau in the same $\sim100\,\rm km/s$ range \cite{yoon2021}, we can expand this distance scale range to a total of 32 orders of magnitude. 

 Another aspect regarding these fundamental speeds is the possible link there is with the radius of the central stars. It is indeed observed that, for many systems, the fundamental speed corresponds approximately to the speed of an object orbiting with a circular trajectory whose radius is the radius $R$ of the star, i.e. $\sqrt{GM/R}$. It was already suggested  that the circular orbit speed at the radius of the Sun could be used as a fundamental speed for a potential intra-mercurial system to predict the presence of very small objects or of structures in the circumsolar dust \cite{nottale2011}. It is in that regard interesting to notice that the main peak of the distribution of fundamental speeds that we found at around $90 \,\rm AU/yr$ ($427 \,\rm km/s$) is very close to the circular speed at the radius of the sun ($437 \,\rm km/s$). While there are no major objects to be found inside of Mercury's orbit in our solar system because of dynamical and thermodynamical constraints, planets in extra-solar systems are very close to their stars. An example of such a system is the three planet system PSR B1257+12, for which the fundamental speed was found to be $426\,\rm km/s$ with great precision \cite{nottale2000}.  Without further interpretation, it is worth noting that the Solar-type stellar radius Keplerian speed ($437 \,\rm km/s$) happens to be very close to twice the super-fundamental speed, $(218.0\pm4.7)\,\rm km/s$, which we identified, and also very close to being the triple of the fundamental speed we found for the inner solar system, $(144\pm3)\,\rm km/s$.

These results are suggestive that the Resolution-Scale Relativity principle would be implemented in planetary systems where it manifests itself via the emergence of macro-quantization. However, the theoretical picture is incomplete. Planets seem to be found at distances from the central stars corresponding to the maxima of the squared radial functions of the $l=n-1$ solutions of a generalized Schr\"odinger equation, in which the Planck constant is replaced with $2m\mathcal D$, where $m$ is the mass of the planet and $\mathcal D$ a diffusion constant, making the de Broglie wavelength independent of the mass $m$. One would instead expect to find them distributed according to the quantum-like $l=n-1$ probability densities for different values of the principal quantum-like number $n$, which strongly overlap each other in such a way that no quantum-like structuring should be observable. This would then imply that some mechanisms unaccounted for in Section \ref{bohr} must be at play to result in the migration of matter towards regions of maximal macro-quantum probability density. These mechanisms could be of dissipative or radiative nature if this migration happens during the proto-planetary phase \cite{gaidos} and would need modeling in the Resolution-Scale Relativity framework.  

Independently from this lack of a theoretical model for the migration or condensation of objects onto classical trajectories corresponding to the maximal quantum-like probability densities, the results presented in this article are encouraging for other similar searches. We can point out some possible investigations focusing on gravitational systems. Concerning exoplanets, all systems with less than four planets (the vast majority) were excluded from the present analysis. These systems require a different approach. Regarding the planetary systems studied in this paper, it would be interesting to study the planets masses with respect to distances in the Resolution-Scale Relativity framework, as it was already done for the Solar System \cite{nottale1997}. While our observations are only indirect evidences of a primordial macro-quantum structure emergence during the planetary formation era, the recent high-resolution observations of proto-planetary disks open the possibility of searches for macro-quantum signatures early on in the evolution of planetary systems.  Moving away from planetary systems, eclipsing binaries, for which orbital parameters are known with great accuracy, could also be studied. These studies are coming in complement to laboratory-based efforts to identify scale relativistic macro-quantization effects \cite{lebohec2017,nottale2024, nottale2024b}.

\section*{Acknowledgement}
This research used of the NASA Exoplanet Archive, which is operated by the California Institute of Technology, under contract with the National Aeronautics and Space Administration under the Exoplanet Exploration Program. The authors are grateful for the comments and suggestions they received from Laurent Nottale, Jamie Holder and Mei-Hui Teh. 

\section*{ORCID}
\noindent Julien Foerster - \url{https://orcid.org/0009-0002-1111-2641}

\noindent Tugdual LeBohec - \url{https://orcid.org/0000-0002-3489-7325}


\begin{thebibliography}{0}

\bibitem{agol2005}E.~Agol et al., "On detecting terrestrial planets with timing of giant planet transits", 2005, MNRAS, 359,  2, 567-579.

\bibitem{bohr1913} Niels Bohr, "On the Constitution of Atoms and Molecules, Part I", Philosophical Magazine,  1913, {\bf 26} (151): 1-24.

\bibitem{caswell1929} A.E.~Caswell, "A relation between the distances of the planets from the Sun", Science. 1929 Apr 5, 69(1788):384.

\bibitem{cohentan}  C. Cohen-Tannoudji, B. Diu, \& F. Lalo\"e, "Quantum Mechanics", Wiley-VCH; 2nd edition (December 4, 2019), ISBN-13  :  978-3527345533
    
\bibitem{gaidos} E.~Gaidos, L.~Gehrig, \& M.~G\"udel, "On the diversification and dissipation of protoplanetary disks", A\&A 696, A207 (2025)

\bibitem{han2014} E.\,Han, S.\,Wang, X.\,Sharon, J.\,Wright, K.\,Feng,  M.\,Zhao, O.\,Fakhouri, J.\,Brown,  C.\,Hancock, "Exoplanet Orbit Database. II. Updates to Exoplanets.org",  Publications of the Astronomical Society of the Pacific, Volume 126, Issue 943, pp. 827 (2014).

\bibitem{hermann1997} R.P.~Hermann, G.~Schumacher \& R.~Guyard, "Scale relativity and quantization of the solar system: Orbit quantization of the planet's satellites", Astron. Astrophys. 335, 281-286 (1998)

\bibitem{lebohec2017} S.~LeBohec, "Scale Relativistic signature in the Brownian motion of micro-spheres in optical traps", Int. J. Mod. Phys. A, 32, 1750156 (2017)

\bibitem{malisoff1929}W.M.~Malisoff, "Some new laws for the solar system", 1929, Science  04 Oct 1929, Vol. 70, Issue 1814, pp. 328-329 

\bibitem{Nelson1966}  E.~Nelson, "Derivation of the Schr\"odinger Equation from Newtonian Mechanics", Phys. Rev. 150, 1079  (1966) 

\bibitem{Nieto1972} M.M.~ Nieto, "The Titius-Bode Law of Planetary Distances: Its History and Theory"; Pergamon Press, 1972; ISBN 0080167845 

\bibitem{nottale1993} L.~Nottale, "Fractal space-time and microphysics"; World Scientific Publishing Company, 1993; ISBN 9810208782

\bibitem{nottale2011} L.~Nottale, "Scale Relativity And Fractal Space-Time: A New Approach to Unifying Relativity and Quantum Mechanics"; Imperial College Press, 2011; ISBN: 978-1-84816-650-9  

\bibitem{nottale1996} L.~Nottale, "Scale-Relativity and Quantization of Extrasolar Planetary Systems", Astron. Astrophys. Lett. 315, L9 (1996).  

\bibitem{nottale1997} L.~Nottale, G.~Schumacher \& J.~Gay, "Scale relativity and quantization of the solar system", Astron. Astrophys. 322, 1018-1025 (1997)

\bibitem{nottale2000} L.~Nottale, G.~Schumacher, \& E.T.~Lef\`evre, "Scale-Relativity and Quantization of Exoplanet Orbital Semi-Major Axes", Astronomy \& Astrophysics, Vol. 361, pp. 379-387 (2000).

\bibitem{nottale2024} L.~Nottale, \& T.~Lehner,  "The turbulent jet in the scale-relativity framework", Physics of Fluids 36(4) DOI: 10.1063/5.0187140

\bibitem{nottale2024b} L.~Nottale, \& T.~Lehner, " Macroscopic quantum-like behavior of a turbulent jet ", 2024, hal-05005616. 

\bibitem{peale76} S. Peale, "Orbital resonance in the solar system", 1976, Annual review of astronomy and astrophysics. Volume 14, 215-246.

\bibitem{prodanov2021} D.~Prodanov, "The Burgers equations and the Born rule", 2021, Chaos, Solitons \& Fractals. Volume 144(4): 110637.

\bibitem{peniston1930} J.B.~Penniston, "A new law of satellite distances", Science  16 May 1930: Vol. 71, Issue 1846, pp. 512-513.

\bibitem{schrodinger1926} E.~Schr\"odinger,  "An Undulatory Theory of the Mechanics of Atoms and Molecules", 1926, Physical Review. {\bf 28} (6): 1049-1070. 	 

\bibitem{mhteh2018} M.-H. Teh, L. Nottale and S. LeBohec, "Scale relativistic formulation of non-differentiable mechanics", Eur. Phys. J. Plus (2019) 134: 438; DOI: 10.1140/epjp/i2019-12840-6

\bibitem{yoon2021} Y.~Yoon et al., "Rotation Curves of Galaxies and Their Dependence on Morphology and Stellar Mass", ApJ, 922:249, 2021
    
\end{thebibliography}
\end{document}